\documentclass[]{spie}  

\usepackage{amsmath,amsfonts,amssymb,aas_macros}
\usepackage{graphicx}
\usepackage[colorlinks=true, allcolors=blue]{hyperref}
\usepackage{enumitem}
\usepackage{makecell}

\title{The Roman Coronagraph Community Participation Program: trials and triumphs of designing an observing program for a technology demonstration instrument}

\author[a]{Schuyler G. Wolff}
\affil[a]{Steward Observatory and the Department of Astronomy, The University of Arizona, 933 N Cherry Ave, Tucson, AZ, 85721, USA}

\authorinfo{Further author information: (Send correspondence to S.G.W.)\\S.G.W.: E-mail: sgwolff@arizona.edu}

\author[b]{Vanessa P. Bailey}
\author[a]{Justin Hom}
\author[c]{Beth Biller}
\author[d]{Dmitry Savransky}
\author[e]{Julien H. Girard}
\author[f]{Ellis Bogat}
\author[g]{\'Oscar Carri\'on-Gonz\'alez}
\author[h]{Clarissa R. Do \'O}
\author[i]{Masayuki Kuzuhara}
\author[j]{Chen Xie}
\author[k]{Maxwell Millar-Blanchaer}
\author[b]{Susan Redmond}
\author[l]{Johan Mazoyer}
\author[g]{Macarena Vega-Pallauta}
\author[g]{Gael Chauvin}
\author[m]{Matthieu Ravet}
\author[n]{Alexandra Greenbaum}
\author[b]{Eric Cady}
\author[a]{Ramya Anche}
\author[n]{James Ingalls}

\author{the Roman Coronagraph Community Participation Program Team}

\affil[b]{Jet Propulsion Laboratory, California Institute of Technology, 4800 Oak Grove Drive, Pasadena CA 91109 USA}
\affil[c]{Institute for Astronomy, University of Edinburgh, EH9 3HJ, Edinburgh UK}
\affil[d]{Sibley School of Mechanical and Aerospace Engineering, Cornell University, Ithaca, NY, 14853, USA}
\affil[e]{Space Telescope Science Institute, 3700 San Martin Drive, Baltimore, MD 21218, USA}
\affil[f]{Laboratoire d'Astrophysique de Marseille, 38 Rue Frédéric Joliot-Curie, 13013 Marseille, France}
\affil[h]{Department of Astronomy, California Institute of Technology, Pasadena, CA 91125, USA}
\affil[g]{Max Planck Institut for Astronomy, K\"onigstuhl 17, 69117 Heidelberg, Germany}
\affil[i]{Astrobiology Center, NINS, 2-21-1 Osawa, Mitaka, Tokyo 181-8588, Japan and National Astronomical Observatory of Japan, NINS, 2-21-1 Osawa, Mitaka, Tokyo 181-8588, Japan}
\affil[j]{William H. Miller III Department of Physics and Astronomy, Johns Hopkins University, 3400 N. Charles Street, Baltimore, MD 21218, USA}
\affil[k]{Department of Physics, University of California, Santa Barbara, CA 93106, USA}
\affil[l]{LIRA, Observatoire de Paris, Université PSL, CNRS, Université Paris Cité, Sorbonne Université, France}
\affil[m]{Université Côte d’Azur, Nice, France}
\affil[n]{IPAC, Mail Code 100-22, Caltech, 1200 E. California Blvd., Pasadena, CA 91125, USA}

\begin{document} 
\maketitle

\begin{abstract}
The Coronagraph Instrument onboard the Nancy Grace Roman Space Telescope serves as a crucial technology pathfinder for the Habitable Worlds Observatory, with on-sky verification of high-contrast imaging techniques and the potential to image a Jupiter analog in reflected light for the first time. Together with the Roman Project Team, the Community Participation Program (CPP) is responsible for target selection, preparatory observations, developing an exposure time calculator, target database, data reduction pipeline, simulation tools, and engagement with the broader community. Here we present an overview of the CPP activities over the past two years with an emphasis on observation planning activities for the initial in-orbit checkout and the first six months of the observation phase. Finally, we present future opportunities for the astronomical community to interact with the data as it becomes public early in the mission.
\end{abstract}

\keywords{Nancy Grace Roman Space Telescope Coronagraph, High Contrast Imaging, Roman Coronagraph Community Participation Program, Target Selection for the Roman Coronagraph}

\section{INTRODUCTION}
\label{sec:intro}  

The Nancy Grace Roman Space Telescope (Roman) is NASA's next astrophysics flagship mission with a launch readiness date of August 30th, 2026. When fully operational in 2027, Roman will provide the astronomical community with an unparalleled opportunity to demonstrate space-based, visible-band coronagraphy with active optics for the first time, using the novel Roman Coronagraph Instrument. The Coronagraph is a technology demonstration designed to prove a variety of flight hardware, software, and observational/data analysis techniques necessary to mature NASA’s ability to build a next-generation exoplanet characterization mission (namely, the Habitable Worlds Observatory: HWO).

In an observation phase (allocated 90 days or 2200 hours) conducted during the first 18 months of the mission, novel starlight suppression technology may enable direct imaging of a Jupiter analog in reflected light and exo-zodiacal light at optical wavelengths for the first time. The Observation Phase represents the only opportunity to test key technologies in space (e.g. performance as a function of stellar magnitude) and quantify sources of uncertainty (e.g. exozodi \cite{2012PASP..124..799R}) during this crucial phase of the design and execution of HWO.

The Community Participation Program (CPP) Team first established by the ROMAN-22 ROSES call is responsible for the execution of the Coronagraph Instrument Observation Phase. Previous SPIE proceedings introduced the structure and goals of the CPP and its working groups \cite{2024SPIE13092E..1IS,2024SPIE13092E..56M, 2024SPIE13092E..55W}. The work described here is primarily focused on the activities of the Observation Planning Working Group over the past two years, but represents a colossal effort by the entire CPP team. By virtue of this being a technology demonstration instrument with finite support, not every coronagraphic mask onboard the Roman Coronagraph is supported for the first semester. Table \ref{tab:modes} lists the required and best-effort modes expected to be commissioned within the first six months of the observation phase.  A complete list of installed masks in available in \citenum{2025JATIS..11b1403R}.  

\begin{table}[b]
\caption{Required and Best Effort Modes for the Roman Coronagraphic Instrument.} 
\label{tab:modes}
\begin{center}   
\renewcommand{\arraystretch}{1.2}
\begin{tabular}{| c | c | c | c | c | c |}
\hline
Band &   Mode & Mask Type & FOV & Coverage & Support \\ \hline \hline

1, 575 nm &   Narrow FoV & Hybrid Lyot & $0.^{\prime\prime}14 - 0.^{\prime\prime}45$ & 360$^{\circ}$ & Required (Imaging), \\
 & Imaging &  & & & Best Effort (Polarimetry) \\ \hline 

1, 575 nm &  Wide FoV & Shaped Pupil & $0.^{\prime\prime}30 - 0.^{\prime\prime}95$ & 360$^{\circ}$ & Best Effort  \\ 
 &  Imaging &  &  &  & (Imaging and polarimetry) \\
\hline

3, 730 nm &  Slit + R$\sim$50  & Shaped Pupil & $0.^{\prime\prime}18 - 0.^{\prime\prime}55$ &  2x65$^{\circ}$ & Best Effort \\ 
 &    Prism Spectroscopy &  &  &  & \\  \hline

4, 825 nm & Wide FoV  & Shaped Pupil & $0.^{\prime\prime}45 - 1.^{\prime\prime}4$ & 360$^{\circ}$ & Best Effort  \\ 
 & Imaging &  & & & (Imaging and polarimetry)  \\ \hline
\end{tabular}
\end{center}
\end{table}

The primary Level-1 requirement of the Roman Coronagraph is to demonstrate a contrast ratio of $10^{-7}$ in the visible at 6 to 9 $\lambda$/D ($0.3^{\prime\prime}-0.45^{\prime\prime}$) in Band 1 (B1, 575 nm) around a bright ($V\sim5$) star (often referred to as TTR5)\cite{2023SPIE12680E..0TB}. This can be accomplished using either the NFOV (Narrow Field of View) and WFOV (Wide Field of View) imaging modes using the HLC or SPC coronagraphs. Then, as time and resources allow, we aim to push performance limits, perform transformative science and maximize the long-term science and technical value to the astronomy community. There will be no exclusive use period for observations obtained by the Roman Coronagraph, and the dataset described in the next sections is meant to serve as a resource for the high-contrast imaging community.

Here we focus on the planned observations for the first six months of Roman Coronagraph operations. Assuming a launch date of August 30th this year, this includes observations spanning mid-December 2026 - June 2027 and represents about a third of the observation phase allocation. We present the methodology for gathering and distilling input from the broader community in Section \ref{sec:input}. The planned target list is presented in Section \ref{sec:targets}. Finally, a brief discussion of future plans and opportunities for community involvement is provided in Section \ref{sec:future}. 

\section{Incorporating Community Input}
\label{sec:input}

\subsection{Community Interest Survey}

The first opportunity for broader input came in the form of a Community Interest Survey in early 2025. The goal of the survey was to conduct a consistency check to confirm the goals of the CPP for the Roman Coronagraph align with the goals of the community.
The survey asked participants to rank various science goals, technology demonstration goals, and observing modes as ``low", ``medium", or ``high" priority. 
Finally, short answer questions allowed participants to elaborate on their top priorities and provide additional suggestions. 

\begin{figure}[!h]
\includegraphics[width=\textwidth]{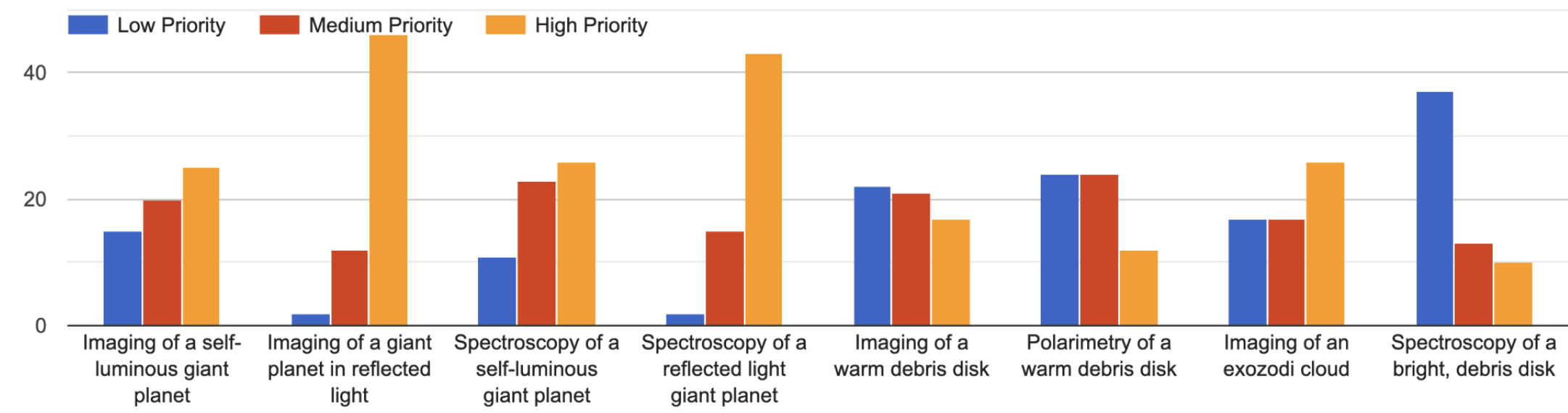}
\includegraphics[width=\textwidth]{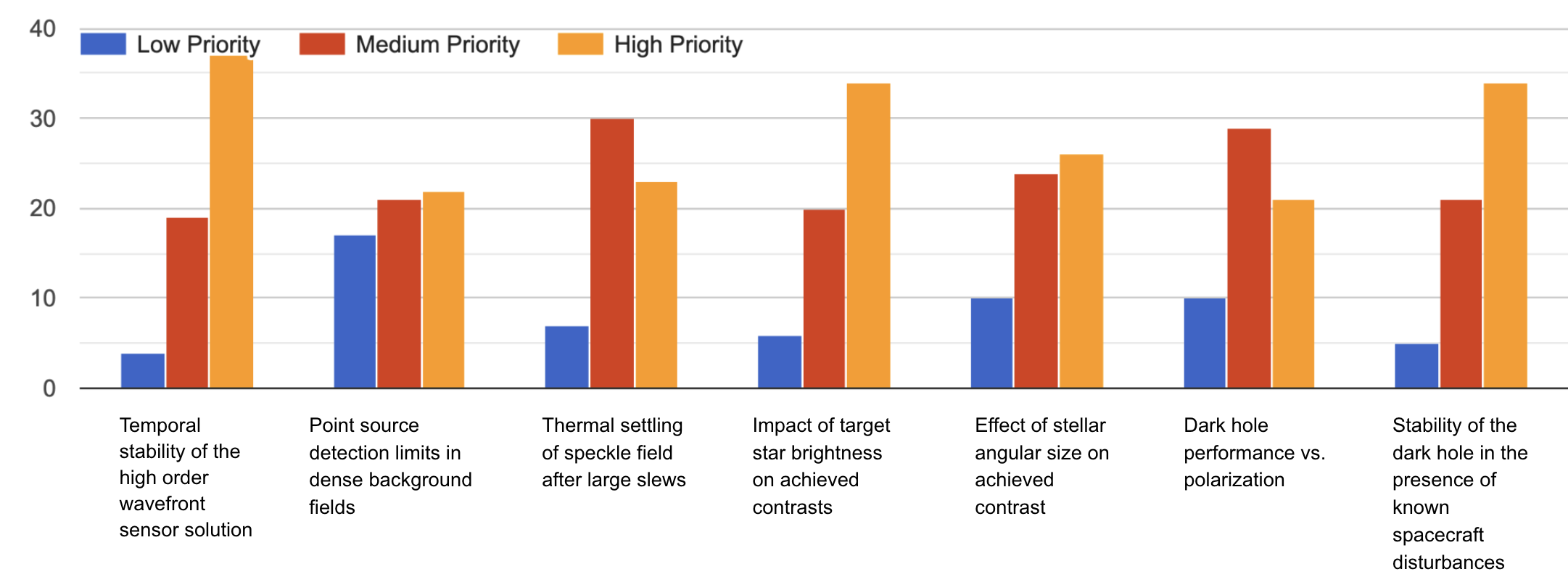}
    \caption{2025 Community Interest Survey results showing the highest priority science (top) and technology demonstration (bottom) activities for the Roman Coronagraph from the 60 survey participants. Dark hole stability and imageing/spectroscopy of a reflected light planet were heavily favored. \label{fig:survey}}
\end{figure}

The survey was introduced during a special AAS Splinter Session in January, 2025 and remained open through March of 2025. To make the survey more accessible, we developed a \href{https://docs.google.com/document/d/1T9aYqe4tgVOAr-Qrg6P-e_bKCa11PPqK0phuPlSe2hE/preview?tab=t.0#heading=h.nj23sjpj5u97}{Roman Coronagraph Primer} to present basic information on the instrument capabilities. The survey received $\sim$ 60 responses; a strong turnout from the high contrast imaging community. The strongest science case preference was for the imaging and spectroscopy of reflected light planets. The highest preferred technology demonstration case was to understand the temporal stability of the dark hole under various performance scenarios. The impact of target star brightness on the achieved contrast was also a high priority choice. Figure \ref{fig:survey} summarizes the results of the survey. 

Survey participants also used the short answer portion to their advantage; providing science and technology cases outside of those specified in the survey and calling out specific high priority science targets. This prompted the CPP to solicit longer form community input with a white paper call detailed in the next section.

\subsection{White Paper Call}

In order to further engage the coronagraph/exoplanet community in the definition of the 90 day observing program, the CPP sponsored a call for white papers to collect concepts for technology demonstration and scientific exploration observations with the Roman Coronagraph in the summer of 2025. 
The white paper template included the observing mode and required contrast, a table of example targets, a description of the scientific and/or technological goals, an observing description, up to 3 supporting figures, and references. Within the observing description section, white paper authors could optionally provide more detail on the required integration times using the CorgiETC\footnote{ \url{https://github.com/roman-corgi/corgietc?tab=readme-ov-file}}. 

In order to maximize participation in the white paper call, the CPP hosted a series of virtual town halls with informational slides and observation planning tool tutorials hosted on our website\footnote{\url{https://www.romancoronagraph.space/}}. A particular effort was made to include Early Career Researchers as both presenters and attendees of the Town Halls. This recruitment effort was surprisingly successful and resulted in a high turnout with over 80 white paper submissions (compare to the 42 white papers submitted to the Strategic Exoplanet Initiatives with HST and JWST that resulted in the Rocky Worlds DDT program \cite{2024arXiv240402932R}) including both community and CPP led efforts. The full list of submitted white papers is available in a Zenodo repository \cite{roman_coronagraph_community_participatio_2026_21203363}. 
Several key themes emerged in the collection:
\setlist{nolistsep}
\begin{itemize}[noitemsep]
    \item Imaging (and polarimetry) of known reflected light and self-luminous planets/substellar companions.
    \item Exozodi imaging: translating from the mid-IR to optical, especially for high-priority HWO targets
    \item Disks: characterization of warmer, transport-dominated, inner components; polarimetry; disk planet interactions etc.
    \item Roman Coronagraph Model validation: Quantify contrast as a function of Target properties (brightness, diameter), DM settling time, thermal drift, Delta pitch between target/reference etc. 
    \item On-sky validations of alternative High Order WaveFront Sensing and Control (HOWFSC) algorithms
\end{itemize}

Several additional science cases were identified but left for future semesters due to feasibility concerns (most related to the Roman Coronagraph $V_{mag} < 7$ requirement) including AGN, evolved stars, and protoplanetary disks. 
This agrees well with the results of the community interest survey and the CPP priorities. While not everything listed above is feasible to include in the first semester, we have endeavored to include as much as possible while balancing the risks and potential rewards. In the following Section, we detail how these white papers were analyzed and incorporated into the Roman Coronagraph observation plan.

\subsection{Prioritizing Observations}

To review all white paper submissions, the CPP formed the Time Prioritization Committee (TPC). The committee consists of 
CPP representatives from US institutions, ESA, CNES, MPIA, and JAXA, and select members from the Roman Coronagraph Instrument Team (at the Jet Propulsion Laboratory, JPL) and the Science Support Center (Infrared Processing and Analysis Center, IPAC). 
In a review process conducted in October of 2025, each white paper  (with authorship information removed) was assessed on either/both science and/or technical merit in addition to feasibility and risk. 
White papers that articulated a path toward transformative advancements for the field, demonstrated that Roman Coronagraph was the only instrument capable of such an advancement, and contained an experimental design that was well-defined and an efficient use of resources were prioritized. 

The outcome of this review process was to establish
overall program themes to be matured by CPP members into Observation Plans. These themes often incorporated multiple elements of related but distinct white paper submissions without fully adapting a single white paper submission into an observing campaign. For example, several white paper submissions focused on observations of different known radial velocity planets for reflected-light imaging; these ideas were agglomerated into observing program themes that aimed to image and characterize reflected-light planet(s) in general. Other programs adopted specific target(s) from white paper(s) but proposed observing them in different instrument configurations not suggested by the white paper itself. 

The Observing Plans were required to provide a much higher level of detail than the white papers with specific target and observation implementation information that was impractical to require of the broader community in the white paper call.  Observing Program authors were required to provide high fidelity exposure time estimates, assess target visibility against the Galactic Bulge Time Domain Survey (along with the limited FOV of the SPC-SPEC Band 3 observing modes, when applicable), incorporate estimates of overheads into total program times, list necessary precursor observations and calibrations, and to briefly describe the analysis approach. 
An oversubscription rate for the first semester of 20\% was chosen to accommodate the commissioning of additional modes and to allow for future updates in our knowledge of the instrument performance and to account for scheduling and planning issues. The resulting ``success-oriented'' target catalog for the first $\sim$ six months of Roman Coronagraph operations in the next section.

\begin{table}[!p]
\caption{Science and technology demonstration targets under consideration for the first six months of the Roman Coronagraph Observation Phase. We provide target IDs, the planned observing mode(s), and a brief description of the primary goals for each observation.} 
\label{tab:targets}
\begin{center}   
\renewcommand{\arraystretch}{1.5}
\begin{tabular}{|p{0.8in}|p{1.4in}|p{3.5in}
|}
\hline
Target ID(s) & Observing Mode(s) & Primary Goal \\
\hline \hline
HIP 71618B & NFOV B1 Imaging \newline B3 Spectroscopy & Exercise imaging and spectroscopy modes on a moderate-contrast companion. \\
\hline
$\beta$ Car & NFOV B1 Imaging & two alternative wavefront sensing-only pilots: using Gaussian probes and polarized imaging.  \\
\hline 
$\epsilon$ Eri b & WFOV B1 Imaging & Constrain albedo and orbit of the known radial velocity (RV) planet, characterize the disk morphology, and search for potential disk/planet interactions. \\
\hline
$\upsilon$ And d 
\newline 14 Her b & NFOV B1 Imaging & Directly image the planet in reflected light; constrain inclination/mass. \\
\hline
$\eta$ Corvi  & NFOV B1 Polarimetry WFOV B1 Polarimetry & Determine if the known exoKuiper belt is the origin of the water- and carbon-rich dust in the habitable zone. \\
\hline 
HR 4796A  & NFOV B1 Polarimetry WFOV B1 Polarimetry WFOV B4 Polarimetry & On-sky validation of all 3 polarimetry modes. NFOV will enable total intensity imaging of the front/backscattering disk regions while the WFOV will probe the disk ansae. \\
  \hline
HD 172555  & NFOV B1 Polarimetry & Edge-on debris disk with an active collisional history. First optical, total-intensity image.  \\
\hline
$\kappa$ And B  \newline HR 2562 B \newline HR 8799 c,d  & WFOV B4 Imaging  & Optical imaging of known self-luminous targets will break degeneracies in atmospheric models (temperature structure, composition, cloud properties). \\
\hline
$\epsilon$ Eri \newline $\beta$ Leo \newline $\beta$ UMa \newline $\tau$ Ceti \newline $\beta$ Vir \newline $\delta$ UMa \newline 72 Her  & NFOV B1 Imaging  & Exozodi pilot program to place meaningful limits on the scattered-light signatures of exozodiacal dust. Targets  include a mix of sources with strong mid-IR zodi detections and solar-type stars to help constrain the impact on future HWO observations.  \\
\hline
HIP 54515 b & B3 Spectroscopy & Test a more challenging spectroscopy target around a fainter host star. \\
\hline
Vega \newline Fomalhaut \newline $\beta$ Leo & WFOV B4 Polarimetry & Investigate the warm, inner debris disk components of nearby stars to probe the impacts of drag forces on the inward dust transport from cold, outer Kuiper-belt analogs. 
\\
\hline
HD 106036 \newline HD 106797 \newline HD 113457 & WFOV B4 Polarimetry & Sco-Cen Debris disk survey to validate ``Star-Hopping'' observing strategy (for targets close in magnitude and on-sky save significant HOWFSC-related overheads). \\
\hline
\end{tabular}
\end{center}
\end{table}

\section{Planned Science and Technology Demonstration Targets for Spring 2027}
\label{sec:targets}

The number one priority for the Roman Coronagraph during early Operations is to demonstrate TTR5, if it is not already achieved during In-Orbit Checkout (commissioning). Once this is complete, we have outlined an ambitious plan to maximize the science and technical return of the Roman Coronagraph in early operations. This includes attempts to directly image as many as three reflected light planets and the goal to commission and exercise all best-effort modes (NFOV B1 imaging and polarization, WFOV B1 and B4 imaging and polarization, and B3 spectroscopy) within the first six months of operations. We summarize the planned targets in Table \ref{tab:targets}. Detailed scheduling and implementation of these observations is still ongoing and much of these plans rely on uncertain information about the on-sky instrument performance (particularly for the goal modes). Thus, this target list  is subject to change. 
Note that Roman targets are required to be single stars with $V_{mag} < 7$ to ensure successful acquisition and limit contrast degradation due to stray light, though this guidance may be updated in future semesters.

There are several classes of observations not included in Table \ref{tab:targets}. We will observe a collection of dark hole digging and PSF reference stars \cite{2026AJ....171...36H} to screen for companions, described in detail in a companion SPIE proceedings \cite{hom2026SPIE}. We will be conducting Low Order Wavefront Sensing and Control (LOWFSC) Acquisition tests on fainter targets with $V_{mag} \sim 5 - 9$. Finally, various calibration targets are also not included in this list. Note also that while many of the high priority technology demonstration goals are not explicitly mentioned, this collection of datasets will provide sufficient diversity in host star magnitudes and diameters, changes in the spacecraft pitch between target and reference star, and thermal environment to provide a rich dataset integrated modeling validation. 

\section{Future Observations and Opportunities}
\label{sec:future}

By the end of the first six months of Roman Coronagraph operations, we may also have enough information to significantly expand the science cases for later on in the Observation phase and a potential extended mission. Early engineering tests will assess if the contrast stability for stars in the 5 - 9 $V_{mag}$ range is acceptable. If confirmed, this would open up exciting new target populations. If we can reach below the bright star limit of GAIA ($V < 6$), more planet candidates might become accessible for reflected-light imaging and/or spectroscopic characterization. Many protoplanetary disks also become accessible in the 7 - 9 $V_{mag}$ range. Investigating these young disk structures and signatures of embedded planets (via SPEC B2 H-$\alpha$ observations) was a popular topic in the 2025 white paper submissions. 
Finally, this would also increase the number of Solar-type stars accessible for an expanded Exozodi Survey \cite{2022PASP..134b4402D}. 

We will also be able to assess critical performance metrics including (but not limited to) the raw contrast floor for a given observing mode, the overheads for high order wavefront sensing and control, the thermal stability of the instrument, and the number of suitable reference stars. All of this information will likely require adjustments to the observation strategy and will dictate how many targets we can fit into the 2200~hr observation allocation.
The CPP aims hold an additional community white paper call in the summer of 2027 which can be informed by these initial findings. 

\newpage 

\acknowledgments 
 This material is based upon work by S.G.W. supported by NASA under award No. 80NSSC24K0217. J. R. H. is funded by NASA under award No. 80NSSC25K0364. 
The research was carried out in part at the Jet Propulsion Laboratory, California Institute of Technology, under a contract with the National Aeronautics and Space Administration (80NM0018D0004). D.S. was supported by NASA under award No. 80NSSC24K0216.

\bibliography{report} 
\bibliographystyle{spiebib} 

\end{document}